\documentclass[intlimits,twoside,a4paper]{article}

\usepackage{amsmath,amssymb}
\usepackage{graphicx}
\usepackage{wrapfig}
\usepackage{siunitx}
\usepackage{cite}

\usepackage[T2A]{fontenc}
\usepackage[cp1251]{inputenc}

\usepackage{cmpj2}

\issue{2012}{15}{4}{43007}

\doinumber{10.5488/CMP.15.43007}

\newcommand{\st}{ \raisebox{.1ex}{${}_{\stackrel{\hbox to 12mm
{\rightarrowfill}}{t\gg\tau_0}}$}} 

\title[Hydrodynamics of phonons]%
{Hydrodynamic states of phonons in insulators%
}

\author[S.A. Sokolovsky]{S.A. Sokolovsky}
\address{Prydniprovs'ka State Academy of  Civil Engineering and Architecture}

\date{Received July 3, 2012, in final form October 1, 2012}
\authorcopyright{S.A. Sokolovsky, 2012}

\begin{document}

\maketitle

\begin{abstract}
The Chapman-Enskog method is generalized for accounting the effect of kinetic modes on hydrodynamic evolution. Hydrodynamic states of phonon system of insulators have been studied in a small drift velocity approximation. For simplicity, the investigation was carried out for crystals of the cubic class symmetry. It has been found that in phonon hydrodynamics, local equilibrium is violated even in the approximation linear in velocity. This is due to the absence of phonon momentum conservation law that leads to a drift velocity relaxation. Phonon hydrodynamic equations which take dissipative processes into account have been obtained. The results were compared with the standard theory based on the local equilibrium validity. Integral equations  have been obtained for calculating the objects of the theory (including viscosity and heat conductivity). It has been shown that in low temperature limit, these equations are solvable by iterations. Steady states of the system have been considered and an expression for steady state heat conductivity has been obtained. It coincides with the famous result by Akhiezer in the leading low temperature approximation. It has been established that temperature distribution in the steady state of insulator satisfies a condition of heat source absence.
\pacs 05.20.Dd, 51.10.+y, 63.20.-e, 63.20.Kr, 66.90.+r, 78.66.Nk.
\end{abstract}

\section{Introduction}

A lot of papers are devoted to investigations of the phonon system of insulators. We mean hydrodynamic states described by densities of additive conserved values or their functions (for a system with broken symmetry, order parameters are added). The description of hydrodynamic processes in insulators is still more complicated due to the presence of Umklapp processes in which the momentum is not conserved. A pioneering research into the role of such processes in the theory of thermal conductivity of insulators belongs to Peierls \cite{Sok_1}. Significant contribution to the study of the problem has been made by Akhiezer \cite{Sok_2} who calculated the thermal conductivity of dielectric crystals in steady states at low tem\-pe\-ra\-tu\-res.

In phonon hydrodynamics, local temperature $T(x,t)$ and drift velocity $u_n(x,t)$ introduced with the help of the local equilibrium distribution, are used instead of the densities of energy $\varepsilon (x,t)$ and momentum $\pi_n(x,t)$ (see, for example, books \cite {Sok_4, Sok_5} and reviews \cite {SSok_3,Sok_6,Sok_7} containing the standard theory). The mentioned works substantiated that the velocity decays with time (slowly at low temperatures) but this relaxation process is not considered to be a manifestation of the kinetic mode of a system. During the recent years, great attention has been paid to the investigation of the  effect of kinetic modes on the system dynamics. Especially the role of the Lviv school on statistical physics (Mryglod, Bryk, Tokarchuk, Omelyan, et al.) should be mentioned. See in this regard, for example, the paper \cite {SSok_2} devoted to the effect of relaxation on hydrodynamic processes in a two-component system, and the review \cite {SSok_1} on the general theory. Similar studies of dielectrics are unknown for the author and it was the motive of the investigation. In the investigation we observed that the local equilibrium distribution at $u_n(x,t)\neq 0$ does not give the leading approximation for the nonequilibrium distribution function. Similar result concerning the violation of the local equilibrium was obtained in our paper \cite{SSok_4} on the polaron kinetics due to the presence of velocity and temperature relaxation in the electron-phonon system. Such a result has not been obtained in \cite{SSok_2} since the dynamics of the system was studied in terms of component energies.

Researchers usually put the Chapman-Enskog method in the basis of hydrodynamic state consideration. In the general formulation of this method, the particle distribution function ${\rm {f}}_{\alpha p}(x, t)$ ($ \alpha $ is the number of species of particles, $ p_n $ is momentum) is considered to be a functional $ {\rm {f}} _ {\alpha p} (x, \xi (t))$ of parameters $ \xi _ \mu (x, t) $ describing the hydrodynamic state (see, for example, \cite{Sok_3}).
Such state is observed after the time $ \tau _0 $ that depends on the initial state of the system and the problem to be solved
\begin{equation}\label{S_1}
{\rm {f}} _ {\alpha p} (x, t)\st{\rm {f}} _ {\alpha p} (x, \xi (t)).
\end{equation}
Relation (\ref{S_1}) is referred to as the functional hypothesis and is a natural assumption allowing one to obtain a closed set of equations for parameters $ \xi _ \mu (x, t) $ and to express all the observed values through these parameters. In the known paper \cite{Sok_2}, Akhiezer actually made the same assumption concerning the structure of the nonequilibrium phonon distribution function in his consideration of stationary states of the phonon system of insulators.

A general study of hydrodynamic states of the phonon system of dielectric crystal is conducted in the standard theory \cite{Sok_4,Sok_5,SSok_3, Sok_6,Sok_7} using the Chapman-Enskog method. Herewith, the coincidence of the distribution function $ {\rm {f}} _ { \alpha p} (x, \xi (t)) $ in the zero approximation in the gradients of hydrodynamic variables ${\rm {f}}_{\alpha p}^{(0)}(x,\xi(t))$ and the Planck distribution with the drift velocity $u_n(x, t)$ and temperature $T(x,t)$
\begin{equation}\label{S_3}
n_{\alpha p} (\xi (x, t)),\qquad n_{\alpha p} (\xi) \equiv \left[\re^{(\varepsilon _ {\alpha p} - p_n u_n)/ T} - 1\right] ^{- 1}	
\end{equation}
is substantiated $[\xi _\mu (x, t)$: $\xi_0(x, t) \equiv T(x, t)$, $\xi_n(x, t) \equiv u_n(x, t)]$. The relation $ {\rm {f}}^{(0)} _{ \alpha p} (x, \xi )=n_{\alpha p}(\xi(x))$ means that the local equilibrium is valid for the system. At low temperatures, this result is grounded with great accuracy by the fact that the number of the Umklapp collisions is small compared to the number of the normal ones. {\it In the present paper, the zero approximation in the gradients ${\rm {f}} _ {\alpha p} ^ {(0)} (x, \xi) $ for the distribution function ${\rm {f}} _ {\alpha p} (x, \xi) $  is calculated not only for low temperatures}. This is done with the appropriate generalization of the Chapman-Enskog method and, in the case of small drift velocity $u_n$, it is shown that ${\rm {f}} _ {\alpha p}^{(0)} (x,\xi) \ne n_{\alpha p}(\xi (x))$ but ${\rm {f}} _ {\alpha p}^{(0)} (x,\xi)|_{u=0}= {\mathop n\limits^o}_{\alpha p}(\xi (x))$ \, (${\mathop n\limits^o}_{\alpha p} (\xi)\equiv [\re^{\varepsilon_{\alpha p} / T}-1]^{-1} $ is the local Planck distribution without velocity). In all the mentioned works \cite{Sok_4,Sok_5,SSok_3, Sok_6,Sok_7}, the drift velocity of phonons $ u_n$ is also considered to be a small value but our consideration yields {\it nonlinear hydrodynamics for a small drift velocity} [it contains nonlinear terms in $u_n(x,t)$]. {\it The established violation of the local equilibrium has principal physical meaning although at low temperatures it is small}.

In the present paper, insulators of a cubic symmetry classes are considered. Even in this case, some crystal quantities have a complicated structure (for example, tensors of the fourth rank). Using a small drift velocity approximation, let us avoid a complicated symmetry research. Among other things, terms in estimations have a complicated tensor structure in all our series expansions in the velocity.
	
In some papers, devoted to phonon hydrodynamics issues, crystalline solids are considered in an isotropic model \cite{Sok_6, Sok_7, SSok_5}. This leads to some simplification of the equations obtained in the present paper that will be analyzed elsewhere. A version of  a $\tau$-approximation for phonon collision integral proposed in \cite{SSok_5} allows one to simplify the equations of the theory too.  Such an approximation is not used in the present paper.  Moreover, a $\tau$-approximation approach can be improved by the method of orthogonal polynomials developed in \cite{Sok_7}.

The outline of the paper is as follows. In {section}~\ref{A}, the basic equations of the theory are presented. In {section}~\ref{B}, the basic equations of the theory are solved in a double perturbation theory in small gradients of hydrodynamic variables and small velocity. As a result, expressions for the fluxes of energy and momentum of the phonons with allowance for dissipative processes are obtained and equations of phonon hydrodynamics are derived. { Section}~\ref{C} discusses the relation of the results with the standard phonon hydrodynamics based on the validity of the local equilibrium. In {section}~\ref{D}, an iteration procedure is built for solving integral equations of the theory in the approximation of low temperatures. In the final {section}~\ref{E}, the steady states of the system are considered.

\section{Basic equations of the theory \label{A}}

A kinetic equation for phonons is put
\begin{equation}\label{S_4}
\frac {{\partial {\rm {f}} _ {\alpha p} (x, t)}} {{\partial t}} = - \frac {{\partial \varepsilon _ {\alpha p}} } {{\partial p_n}} \frac {{\partial {\rm {f}}_{\alpha p} (x, t)}} {{\partial x_n}} + I_ {\alpha p} ({\rm {f}} (x, t))	
\end{equation}
in the basis of the consideration.
The main contribution to the collision integral corresponds to three-phonon processes
\begin{eqnarray}
I_ {\alpha p} ({\rm {f}}) &=& \sum_ {\alpha _1 \alpha _2 \alpha _3} {\int_B {\rd ^ 3 p_1 \rd ^ 3 p_2 \rd ^ 3 p_3} } | \Phi (12,3) | ^ 2 \delta (\varepsilon _1 + \varepsilon _2 - \varepsilon _3) \sum_n {\delta (p_1 + p_2 - p_3 - \hbar b_n) }
\nonumber\\
&&{}
\times \left[\delta _ {\alpha \alpha _3} \delta (p - p_3) - \delta _ {\alpha \alpha _1} \delta (p - p_1) - \delta _ {\alpha \alpha _2} \delta (p - p_2) \right] \big[{\rm {f}} _1 {\rm {f}} _2 (1 + {\rm {f}} _3) - (1 + {\rm {f}}_1) (1 + {\rm {f}} _2) {\rm {f}} _3 \big]	 \label{S_5}
\end{eqnarray}
[abbreviations of the type $\varepsilon_{\alpha _i p_i} \equiv \varepsilon _i$, ${\rm {f}} _ {\alpha _i p_i} \equiv {\rm {f}} _i$, $\Phi (\alpha _1 p_1; \alpha _2 p_2, \alpha _3 p_3) \equiv \Phi (12,3)$ are used; $\alpha$, $p_n$ are polarization and quasi-momentum of a phonon; the integrals are taken over the basic cell of a reciprocal lattice $B$; for simplicity,  the term ``momentum'' is used instead of ``quasi-momentum'']. The collision integral is divided into the contribution of normal processes and Umklapp ones $I_ {\alpha p} ({\rm {f}}) = I_{\alpha p}^ N({\rm {f}})+I_{\alpha p}^U ({\rm {f)}}$ in the usual manner [in \eqref{S_5} summand with $n=0$ gives $I_{\alpha p}^ N({\rm {f}})$]. The hydrodynamic equations are a consequence of the phonon energy conservation law and the law of system momentum change
\begin{equation}\label{S_6}
\frac {{\partial \varepsilon (x, t)}} {{\partial t}} = - \frac {{\partial q_l (x, t)}} {{\partial x_l}}\,,\qquad	\frac {{\partial \pi _n (x, t)}} {{\partial t}} = - \frac {{\partial t_ {nl} (x, t)}} {{\partial x_l}} + R_n (x, t).
\end{equation}
Densities of energy and momentum  $\varepsilon (x, t)$, $\pi _n (x, t)$ of the phonon system, corresponding flux densities $q_l (x, t)$, $t_ {nl} (x, t)$, and that of a frictional force $R_n (x, t)$ are given by formulas
\begin{eqnarray}
\varepsilon (x, t) &\equiv& \sum_ \alpha {\int_B {\rd \tau _p \varepsilon _{\alpha p} {\rm {f}} _ {\alpha p} (x, t )}},\qquad
\pi _n (x, t) \equiv \sum_\alpha {\int_B {\rd \tau _p p_n {\rm {f}} _ {\alpha p} (x, t)}},
\nonumber\\
q_l (x, t) &\equiv& \sum_ \alpha {\int_B {\rd \tau _p \varepsilon _ {\alpha p} \frac {{\partial \varepsilon _ {\alpha p}}} {{ \partial p_l}} {\rm {f}} _ {\alpha p} (x, t)}},\qquad t_ {nl} (x, t) \equiv \sum_ \alpha {\int_B {\rd \tau _p p_n \frac {{\partial \varepsilon_ {\alpha {\rm {p}}}}} {{\partial p_l}} {\rm {f}}_{\alpha p} (x, t)}},
\nonumber\\
R_n (x, t) &\equiv& \sum_ \alpha {\int_B {\rd \tau _p p_n I_ {\alpha p} ^ U ({\rm {f}} (x, t))}},\qquad \left[\rd \tau _p \equiv \frac {1} {{(2 \pi \hbar) ^ 3}} \rd ^ 3 p\right].
\label{S_7}
\end{eqnarray}

Temperature $T(x,t)$ and phonon drift velocity $u_n(x, t)$ [variables $\xi _\mu (x, t)$]  are used in this paper instead of densities $\varepsilon(x, t)$, $\pi_n(x, t)$ as independent variables in hydrodynamics. In the standard theory \cite{Sok_4,Sok_5,SSok_3, Sok_6,Sok_7}, the densities are expressed through $\xi _\mu (x, t)$ with the definitions
\begin{equation}\label{S_8}
\varepsilon (\xi)=\sum_\alpha \int_B \rd \tau _p \varepsilon_{\alpha p} n_ {\alpha p} (\xi),\qquad \pi_n(\xi)=\sum_ \alpha \int_B \rd \tau _p p_n n_ {\alpha p} (\xi)
\end{equation}
where $n_ {\alpha p} (\xi)$ is the Planck distribution  (\ref{S_3}). Below, we use the details connected with these relations only while comparing the presented theory with the standard theory according to the above remarks of ours on the violation of the local equilibrium for $u_n\neq 0$.

Investigating fundamental issues of phonon kinetics, we restrict our consideration to the crystals of cubic symmetry classes $O$, $O_h$, $T_h$ for simplicity. In the present paper, the phonon drift velocity $u_n$ is assumed to be small compared to the second sound velocity (let $\lambda$ be a corresponding small parameter). In this case, the energy and momentum densities of the phonon gas can be expanded in powers of $u_n$ as it follows
\begin{equation}\label{S_14}
\varepsilon (\xi) \equiv \varepsilon (T, u) = {\mathop\varepsilon\limits^o} (T) + a (T) u ^ 2 + O \left(\lambda ^ 4\right), \qquad \pi _n (\xi) \equiv \pi _n (T, u) = \sigma (T) u_n + O \left(\lambda ^ 3\right)	
\end{equation}
where functions ${\mathop\varepsilon\limits^o}(T)$, $\sigma (T)$, $a (T)$ are considered to be known. In the standard theory \cite{Sok_4,Sok_5,SSok_3, Sok_6,Sok_7}, they are calculated from (\ref{S_8}). The terms given in (\ref{S_14}) by estimations have a complicated tensor structure even for the mentioned symmetry classes. Their complete investigation can be made using the ideas developed, for example, in \cite{Sok_8}.

Hydrodynamic equations for the variables $\xi _ \mu (x, t)$ have the structure
\begin{equation}\label{S_9}
\frac {\partial \xi _ \mu (x, t)}{\partial t} = M_\mu\big(x,{\rm {f}}(\xi (t))\big),\qquad \left[\frac {\partial^s \xi_\mu (x, t)} {\partial x_{n_1}\dots \partial x_{n_s}}\sim g^s, \qquad g\ll 1\right]	
\end{equation}
and, as usual, the gradients of these variables are small [in (\ref{S_9}) $g=l / L $, where $ l $ is a mean free path of particles, $ L $ is a characteristic size of inhomogeneities in the system]. Functions $M_ \mu (x, {\rm {f}})$ can be expressed through the distribution function ${\rm {f}} _ {\alpha p} (x, \xi)$ with the help of relations (\ref{S_6}), (\ref{S_7}), and (\ref{S_14}). According to (\ref{S_1}) and (\ref{S_9}), the functional ${\rm {f}} _ {\alpha p} (x, \xi)$ satisfies the integro-differential equation
\begin{equation}\label{S_10}
\sum_\mu {\int {\rd ^ 3 x '\frac {{\delta {\rm {f}} _ {\alpha p} (x, \xi)}} {{\delta \xi _ \mu (x ')}}}} M_ \mu (x', {\rm {f}} (\xi)) = - \frac {{\partial \varepsilon _ {\alpha p}}} {{\partial p_n}} \frac {{\partial {\rm {f}} _ {\alpha p} (x, \xi)}} {{\partial x_n}} + I_ {\alpha p} ({\rm {f }} (x, \xi))
\end{equation}
and additional conditions
\begin{equation}\label{S_11}
\sum_\alpha {\int_B {\rd \tau _p \varepsilon _ {\alpha p} {\rm {f}} _ {\alpha p} (x, \xi)}} = \varepsilon (\xi (x)), \qquad \sum_\alpha {\int_B {\rd \tau _p p_n {\rm {f}}_ {\alpha p} (x, \xi) = \pi _n (\xi (x))}}
\end{equation}
that define the phonon gas temperature and drift velocity together with (\ref{S_14}).

\section{Phonon hydrodynamics \label{B}}

Equations (\ref{S_10}), (\ref{S_11}) are solved by us in the double perturbation theory in gradients (the small parameter $g$) and in the drift velocity (the small parameter $\lambda$)
\begin{equation}\label{S_15}
{\rm {f}} _ {\alpha p} (x, \xi) = {\rm {f}} _ {\alpha p} ^ {(0)} + {\rm {f}} _ {\alpha p} ^ {(1)} + O \left(g ^ 2\right),\qquad {\rm {f}} _ {\alpha p} ^ {(s)} = {\rm {f}} _ {\alpha p} ^ {(s, 0)} + {\rm {f}} _ {\alpha p} ^ {(s, 1)} + O \left(g ^ s \lambda ^ 2\right);\qquad 	{\rm {f}} _ {\alpha {\rm {p}}} ^ {{\rm {(0}} {\rm {, 0)}}} = {\mathop n\limits^o}_{\alpha p}\,.
\end{equation}
The Planck distribution without velocity ${\mathop n\limits^o}_{\alpha p}$ is the main contribution to the phonon distribution function. Our calculation is based on the estimations
\begin{equation}\label{S_16}
T \sim \lambda ^ 0 g ^ 0, \qquad u_n \sim \lambda ^ 1 g ^ 0, \qquad \frac {{\partial T}} {{\partial x_n}} \sim\lambda^0g^1, \qquad \frac {{\partial u_n}}{{\partial x_l}}\sim g^1\lambda^1.  	
\end{equation}
The simplest and the most important contributions to the theory have the following structure
\[
{\rm {f}} _ {\alpha p} ^ {(0,1)} = {\mathop n\limits^o}_{\alpha p} \left(1 + {\mathop n\limits^o}_{\alpha p }\right) A_ {\alpha n} (p) u_n\,, \qquad
{\rm {f}} _ {\alpha {\rm {p}}} ^ {{\rm {(0}} {\rm {, 2)}}} = {\mathop n\limits^o}_{\alpha p} \left(1 + {\mathop n\limits^o}_{\alpha p}\right) B_{\alpha nl} (p) u_n u_l\,,	
\]
\begin{equation}\label{S_18}
{\rm {f}} _ {\alpha {\rm {p}}} ^ {{\rm {(1}} {\rm {, 0)}}} = {\mathop n\limits^o}_{\alpha p} \left(1 + {\mathop n\limits^o}_{\alpha p}\right) C_ {\alpha n} (p) \frac {{\partial T}} {{\partial x_n}}\,,
\qquad {\rm {f}} _ {\alpha {\rm {p}}} ^ {{\rm {(1}} {\rm {, 1)}}} = {\mathop n\limits^o}_{\alpha p} \left(1 + {\mathop n\limits^o}_ {\alpha p}\right) \left[D_ {\alpha nl} (p) \frac {{\partial u_n}} {{\partial x_l}} + E_ {\alpha nl} (p) \frac {{\partial T}} {{\partial x_n}} u_l \right].	
\end{equation}
These formulas contain functions $A_ {\alpha n} (p)$, $C_ {\alpha n} (p)$, $D_ {\alpha nl} (p)$ that satisfy equations
\begin{equation}\label{S_19}
\frac {\nu} {\sigma} A_ {\alpha n} (p) = \sum \limits_ {\alpha '} {\int \limits_B {\rd ^ 3 p'} K_ {\alpha \alpha '}} \left(p, p '\right) A_ {\alpha' n} (p '),
\qquad \langle p_n A_ {\alpha n} (p) \rangle = 3 \sigma;
\end{equation}
\begin{equation}\label{S_21}
A_ {\alpha n} (p) \frac {1} {\sigma} \left(\nu _T + \frac {{\partial p}} {{\partial T}}\right) - \frac {1} {{T ^ 2}} \varepsilon _ {\alpha p} \frac {{\partial \varepsilon _ {\alpha p}}} {{\partial p_n}}= \sum_ {\alpha '} {\int \limits_B {\rd ^ 3 p '} K_ {\alpha \alpha'}} \left(p, p '\right) C_ {\alpha' n} \left(p '\right),
\qquad \langle p_n C_ {\alpha n} (p) \rangle = 0;
\end{equation}
\[
\frac {h} {{cT ^ 2}} \varepsilon _ {\alpha p} \delta _ {nl} + \frac {\nu} {\sigma} D_ {\alpha nl} (p) - \frac {{\partial \varepsilon _ {\alpha p}}} {{\partial p_l}} A_ {\alpha n} (p) = \sum \limits_ {\alpha '} {\int \limits_B {\rd ^ 3 p '} K_ {\alpha \alpha'}} \left(p, p '\right) D_ {\alpha' nl} \left(p '\right),
\]
\begin{equation}\label{S_22}
\langle \varepsilon _ {\alpha p} D_ {\alpha nn} (p) \rangle = 0.
\end{equation}
Here, heat capacity $c$, pressure $p$ of the phonon gas and a special average $\langle g_ {\alpha p} \rangle$
\begin{equation}\label{S_23}
c = \frac {{\partial {\mathop\varepsilon\limits^o}}} {{\partial T}} = \frac{\langle \varepsilon _ {\alpha p} ^ 2 \rangle }{ T ^ 2}\,, \qquad
p = \frac {1} {3} \int {\rd \tau _p} p_n \frac {{\partial \varepsilon _ {\alpha p}}} {{\partial p_n}} {\mathop n\limits^o}_{\alpha p}\,;\qquad
\langle g_ {\alpha p} \rangle \equiv \sum_\alpha {\int_B {\rd \tau _p {\mathop n\limits^o}_{\alpha p} \left(1 + {\mathop n\limits^o}_{\alpha p}\right) g_ {\alpha p}}}	
\end{equation}
are introduced for an arbitrary function $g_ {\alpha p}$. Equations for $B_{\alpha nl}(p)$ and $E_ {\alpha nl} (p)$  are not written due to their complexity. The kernel $K_ {\alpha \alpha '} (p, p')$ of the integral equations (\ref{S_19})--(\ref{S_22}) is defined by relations
\[
I_ {\alpha p} \left({\mathop n\limits^o} + \delta {\rm {f}}\right) = \sum_ {\alpha '} {\int {\rd ^ 3 p'M_ {\alpha \alpha '} (p, p')}} \delta {\rm {f}} _ {\alpha 'p'} + O\left(\delta {\rm {f}}^2\right),
\]
\begin{equation}\label{S_24}
{\mathop n\limits^o}_{\alpha p} \left(1 + {\mathop n\limits^o}_{\alpha p}\right) K_ {\alpha \alpha '} \left(p, p'\right) = - M_ {\alpha \alpha '} \left(p, p'\right) {\mathop n\limits^o}_ {\alpha 'p'} \left(1 + {\mathop n\limits^o}_ {\alpha 'p'}\right).
\end{equation}

Equations (\ref{S_19})--(\ref{S_22}) contain the values $a$, $\sigma$ defined by (\ref{S_14}) and depending on $T$. The meaning of variables $\nu$, $\nu_T$, $h$ (they are also functions of the temperature) in these equations becomes clear from the following expressions for the fluxes of momentum, energy, and frictional force density
\[
t_ {n \, l} = p \delta _ {nl} + \mu _ {nl, ms} u_m u_s - \eta _ {nl, ms} \frac {{\partial u_m}} {{\partial x_s} } - \alpha _ {nl, ms} u_s \frac {{\partial T}} {{\partial x_m}} + O \left(g ^ 0 \lambda ^ 3, g ^ 1 \lambda ^ 2\right),
\]
\begin{equation}\label{S_26}
q_l = hu_l - \kappa \frac {{\partial T}} {{\partial x_l}} + O \left(g ^ 0 \lambda ^ 3, g ^ 1 \lambda ^ 2\right),\qquad R_l = - \nu u_l - \nu _T \frac {{\partial T}} {{\partial x_l}} + R_l ^ {(2,1)} + O \left(g ^ 0 \lambda ^ 3, g ^ 1 \lambda ^ 2, g ^ 2 \lambda ^ 2\right).	
\end{equation}
These expressions are obtained in the perturbation theory from the definitions (\ref{S_7}) taking into account the contributions (\ref{S_18}) to the phonon distribution function ${\rm {f}}_{\alpha p} (x, \xi)$. The term $R_l^{(2,1)}$ is included in (\ref{S_26}) because it gives  contribution of the same order as a viscous momentum flux $t^{(1,1)}_{nl}$ in hydrodynamic equation for the velocity [see (\ref{S_31})]. Contributions ${\rm {f}} _ {\alpha p} ^ {(2,0)}$, ${\rm {f}} _ {\alpha p} ^ {(2,1)}$ to the function ${\rm {f}} _ {\alpha p} (x, \xi)$ are required in order to calculate $R_l^ {(2,1)}$  (they will be analyzed in another paper). In (\ref{S_26}) $\nu$, $\nu _T$ are damping rates, $\kappa$ is a thermal conductivity, $\eta_{nl, ms}
$ is a viscosity tensor, $h$ is a coefficient of the drift energy transfer, $
\mu_{nl, ms}$ is a coefficient of the drift momentum transfer, $\alpha_{nl, ms}$ is a coefficient of convective momentum transfer. Their values are given by formulas
\begin{eqnarray}
h &=& \frac {1} {3} \left\langle \varepsilon _ {\alpha p} \frac {{\partial \varepsilon _ {\alpha p}}} {{\partial p_n}} A_ {\alpha n} ( p) \right\rangle,\qquad
\kappa = - \frac {1} {3} \left\langle \varepsilon _ {\alpha p} \frac {{\partial \varepsilon _ {\alpha p}}} {{\partial p_n}} C_ {\alpha n } (p) \right\rangle,
\nonumber\\
\eta _ {nl, ms} &=& - \left\langle p_n \frac {{\partial \varepsilon _ {\alpha p}}} {{\partial p_l}} D_ {\alpha ms} (p) \right\rangle,\qquad
\mu _ {nl, ms} = \left\langle p_n \frac {{\partial \varepsilon_{\alpha p}}} {{\partial p_l}} B_ {\alpha ms} (p) \right\rangle,
\nonumber\\
\alpha_{nl, ms}&=&-\left\langle p_n \frac{{\partial\varepsilon_{\alpha p}}} {{\partial p_l}} E_{\alpha ms} (p) \right\rangle; \qquad
\nu = \frac {1} {3} \left\{p_n, A_ {\alpha n} (p) \right\} ^ U, \qquad
\nu _T = \frac {1} {3} \left\{p_n, C_ {\alpha n} (p) \right\} ^ U.	
\label{S_27}
\end{eqnarray}
Here, damping rates are written in terms of a bilinear form defined by the relation
\begin{eqnarray}
\left\{g_ {\alpha p}, h_ {\alpha p} \right\} ^ {N, U}&=& -\frac {1} {{(2 \pi \hbar) ^ 3}} \sum_ {\alpha \alpha '} {\int_B {\rd ^ 3 p\rd ^ 3 p'g_ {\alpha p} M_ {\alpha \alpha'} ^ {N, U} \left(p, p '\right) {\mathop n\limits^o} _ {\alpha 'p'} \left(1 + {\mathop n\limits^o} _ {\alpha 'p'}\right) h_ {\alpha 'p'}}}\,,
\nonumber\\
\left\{g_ {\alpha p}, h_ {\alpha p} \right\} &=& \left\{g_ {\alpha p}, h_ {\alpha p} \right\} ^N+\left\{g_ {\alpha p}, h_ {\alpha p} \right\}^U
\label{S_28}
\end{eqnarray}
where $M_ {\alpha\alpha'}^ N(p, p')$, $M_{\alpha\alpha'}^U(p, p')$ are contributions to the kernel $M_ {\alpha \alpha '} (p, p')$
associated with the normal and Umklapp processes, correspondingly. Forms of this type are very useful in the kinetic theory and have the properties
\begin{equation}\label{S_29}
\left\{g_ {\alpha p}, h_ {\alpha p} \right\} ^ {N, U} = \left\{h_ {\alpha p}, g_ {\alpha p} \right\} ^ {N, U}, \qquad \left\{g_ {\alpha p}, g_ {\alpha p} \right\} ^ {N, U} \geqslant  0
\end{equation}
that can be proved through the usual way for arbitrary functions $g_ {\alpha p}$, $h_ {\alpha p}$ (see, for example, \cite{Sok_4}).

Expressions (\ref{S_27}) for $\nu$ and $\nu_T$ follow from the integral equations (\ref{S_19}), (\ref{S_21}) taking into account the additional condition for $A_ {\alpha n}(p)$ from (\ref{S_19}) and expression (\ref{S_23}) for pressure $p$. Formula (\ref{S_27}) for $h$ follows from equations (\ref{S_22}) taking into account the definition of the heat capacity $c$ (\ref{S_23}) and the symmetry of the bilinear form $\{g_ {\alpha p}, h_ {\alpha p} \}$ (\ref{S_29}). Therefore, {\it equations} (\ref{S_19})--(\ref{S_22}) {\it can be solved without taking into account the expressions} (\ref{S_27}) {\it for $\nu$, $\nu_T$ and $h$}.

The equations of phonon hydrodynamics (\ref{S_9}) can be written in the form
\begin{eqnarray}
c \frac {{\partial T}} {{\partial t}} &=& \left \{{- h \frac {{\partial u_n}} {{\partial x_n}}} \right \} - \frac { {\partial h}} {{\partial T}} \frac {{\partial T}} {{\partial x_n}} u_n + \frac {{2av}} {\sigma} u_n \left [{u_n + \frac {1} {\nu} \left ({\nu _T + \frac {{\partial p}} {{\partial T}}} \right) \frac {{\partial T}} {{\partial x_n }}} \right] - \frac {{\partial q_n ^ {(1,0)}}} {{\partial x_n}} + O \left(g ^ 0 \lambda ^ 3, g ^ 1 \lambda ^ 2, g ^ 2 \lambda ^ 2\right),
\nonumber\\
\sigma \frac {{\partial u_n}} {{\partial t}} &=& \left \{{- \nu u_n - \frac {{\partial p}} {{\partial T}} \frac {{\partial T}} {{\partial x_n}}} \right \} - \nu _T \frac {{\partial T}} {{\partial x_n}} - \frac {{\partial t_ {nl} ^ {( 1,1)}}} {{\partial x_l}} + R_n ^ {(2,1)} - \frac {1} {c} \frac {{\partial \sigma}} {{\partial T}} u_n \frac {{\partial q_l ^ {(1,0)}}} {{\partial x_l}} + O \left(g ^ 0 \lambda ^ 3, g ^ 1 \lambda ^ 2, g ^ 2 \lambda ^ 3\right)
\label{S_31}
\end{eqnarray}
with accounting relations (\ref{S_6}), (\ref{S_7}), (\ref{S_15}), (\ref{S_18}). Only the terms given in curly brackets were obtained, for example, in book \cite{Sok_4} that discusses the standard theory [in the standard theory, $h$ and $\nu$ should be taken from (\ref{S_34})]. Other terms in \eqref{S_31} are obtained in the present paper (the corresponding contributions of the standard theory were not found in \cite{Sok_4}). Dissipative fluxes $q_n^{(1,0)}$, $t_{nl}^{(1,1)}$  are given in (\ref{S_26}). {\it The obtained equations of the phonon hydrodynamics} \eqref{S_31} {\it fully take into account all the dissipative processes}.

The integral equation (\ref{S_19}) is an equation for eigenfunction $A_{\alpha n}(p)$ and eigenvalue $\nu / \sigma$. {\it The solution of this spectral problem describes the kinetic mode announced above}.  According to equation \eqref{S_31} the velocity really decays  if the ratio $\nu/\sigma$ is positive. Due to (\ref{S_29}), the positivity of the coefficient $\nu/\sigma$  follows from the relation
\begin{equation}\label{S_32}
\frac {\nu} {\sigma} \left\langle A_ {\alpha n} (p) ^ 2 \right\rangle = \left\{A_ {\alpha n} (p), A_ {\alpha n} (p) \right\}
\end{equation}
that can be derived from equation (\ref{S_19}) and definitions (\ref{S_23}) and (\ref{S_28}). {\it This result is obtained in the framework of the considered theory which implies that the phonon drift velocity is small and has no obvious restrictions for the temperature}.

The considered hydrodynamic states occur when $t \gg \tau _0$ where time $\tau _0$, introduced by the functional hypothesis \eqref{S_1}, should be much smaller than the drift velocity decay time $\tau _u$ [$\tau _u = \sigma / \nu$ due to (\ref{S_31})].

\section{Comparison with the standard phonon hydrodynamics \label{C}}

In the standard theory \cite{Sok_4,Sok_5,SSok_3, Sok_6,Sok_7}, the hydrodynamic distribution function in the zero order approximation in gradients ${\rm {f}} _ {\alpha p} ^ {(0)} (x, \xi)$ coincides with the local Planck distribution with velocity $n_ {\alpha p} (\xi (x))$. The functions $A_ {\alpha n} (p)$, $B_ {\alpha nl} (p)$ according to (\ref{S_3}), (\ref{S_15}), and (\ref{S_18}) are given by expressions
\begin{equation}\label{S_33}
A_ {\alpha n} ^ o (p) = \frac{1}{T}p_n\,, \qquad 	B_ {\alpha nl} ^ o (p) =
\frac{1}{2T^2}\left(1 + 2 {\mathop n\limits^o}_{\alpha p}\right) p_n p_l\,.
\end{equation}
These functions are not solutions of the integral equation (\ref{S_19}) and the corresponding equation for $B_ {\alpha nl}(p)$, i.e. $A_{\alpha n}^o(p)\neq A_{\alpha n}(p)$, $B_ {\alpha nl}^o(p)\neq B_{\alpha nl}(p)$. The first inequality means that {\it even in the linear approximation in drift velocity,  ${\rm {f}} _ {\alpha p} ^ {(0)} (x, \xi) \ne$ $n_ {\alpha p} (\xi (x))$ and, therefore, the local equilibrium is violated in the phonon hydrodynamics}. The result $A_{\alpha n}^o(p)\neq A_{\alpha n}(p)$, $B_ {\alpha nl}^o(p)\neq B_{\alpha nl}(p)$ follows from the next section where it is shown that expressions (\ref{S_33}) give only the leading contribution to $A_{\alpha n}(p)$, $B_{\alpha nl}(p)$ in the low temperature approximation.

In the standard theory, temperature and drift velocity are defined by formulas (\ref{S_8}) giving the following expressions for values $\sigma$ and $a$ in (\ref{S_14})
\begin{equation}\label{S_01}
\sigma^o=\frac{1}{3T}\left\langle p^2\right\rangle, \qquad a^o=\frac{1}{6T^2}\left\langle \varepsilon_{\alpha p}(1+2{\mathop n\limits^o}_{\alpha p})p^2\right\rangle
\end{equation}
(hereinafter we denote the value $A$ of the standard theory by  $A^o$).

Formulas (\ref{S_27}) can be also obtained in the standard approach but not all corresponding contributions to phonon hydrodynamic equations are discussed in \cite{Sok_4,Sok_5,SSok_3, Sok_6,Sok_7}. Therefore, (\ref{S_27}) and (\ref{S_33}) lead to expressions for the damping rates $\nu$, $\nu_T$ and drift transfer coefficients $h$, $\mu_{nl,ms}$
\[
\nu ^ o = \frac {1} {{3T}} \left\{p_n, p_n \right\} ^ U,
\qquad \nu ^ o_T = \frac {1} {3T} \left\{p_n, C^o_{\nu n} \right\} ^ U, \qquad
h^o = \frac {1} {{3T}} \left\langle \varepsilon _ {\alpha p} \frac {{\partial \varepsilon _ {\alpha p}}} {{\partial p_n}} p_n \right\rangle = T \frac {{\partial p}} {{\partial T}}= {\mathop\varepsilon\limits^o} + p,
\]
\begin{equation}\label{S_34}
\mu _ {nl, ms} ^ o = \frac {1} {{2T ^ 2}} \left\langle p_n \frac {{\partial \varepsilon _ {\alpha p}}} {{\partial p_l}} p_m p_s \left(1 + 2 {\mathop n\limits^o}_{\alpha p}\right) \right\rangle.	
\end{equation}
Equations (\ref{S_21})--(\ref{S_22}) for the functions $C_ {\alpha n} (p)$ and $D_ {\alpha nl} (p)$ are simplified by the first formula in (\ref{S_33}). Therefore, according to equation (\ref{S_21}) and expression (\ref{S_27}) the phonon thermal conductivity is given in the standard theory by formula
\begin{equation}\label{S_35}
\kappa ^ o= \frac {{T ^ 2}} {3} \left\{C_ {\alpha n} ^ o (p), C_ {\alpha n} ^ o (p) \right\}.
\end{equation}

\section{Solution of the integral equations of the theory at low temperatures \label{D}}

At low temperatures $T \ll T_{\mathrm{D}}$, the following estimates of the kernels of integral equations (\ref{S_19})--(\ref{S_22}) are valid
\begin{equation}\label{S_36}
K_ {\alpha \alpha '} ^ N \left(p, p'\right) \sim \mu ^ 0,\qquad
K_ {\alpha \alpha '} ^ U \left(p, p'\right) \sim \mu ^ 1,\qquad \left(\mu \equiv \re ^ {- T_{\mathrm{D}} / T}\right)
\end{equation}
(see, for example \cite{Sok_2,Sok_4}) where $T_{\mathrm{D}}$ is Debye temperature which is equal to the maximal energy of a phonon.

The solutions of equations (\ref{S_19})--(\ref{S_22}) for functions $A_ {\alpha n}(p)$, $C_ {\alpha n}(p)$, and $D_ {\alpha nl}(p)$ and the corresponding equation for $B_ {\alpha nl}(p)$, damping rates $\nu, \nu_T$ are found in the form
\begin{eqnarray}
A_ {\alpha n} &=& A_ {\alpha n} ^ {[0]} + A_ {\alpha n} ^ {[1]} + O \left(\mu ^ 2\right), \qquad \nu = \nu ^ {[1]} + \nu ^ {[2]} + O \left(\mu ^ 3\right);\qquad B_ {\alpha nl} = B_{\alpha nl}^{[0]} + B_{\alpha nl}^{[1]} + O\left(\mu ^ 2\right);
\nonumber\\
C_ {\alpha n} &=& C_ {\alpha n} ^ {[0]} + C_ {\alpha n} ^ {[1]} + O \left(\mu ^ 2\right),\qquad
\nu _T = \nu _T ^ {[1]} + \nu _T ^ {[2]} + O \left(\mu ^ 3\right); \qquad D_ {\alpha nl} = D_{\alpha nl}^{[0]} + D_{\alpha nl}^{[1]} + O\left(\mu ^ 2\right)
\label{S_37}
\end{eqnarray}
($A ^{[s]}$ is a contribution of the order $\mu ^ s$  to the quantity $A$).

It is easily to understand that {\it the mentioned main equations of the developed theory} (\ref{S_19})--(\ref{S_22}) {\it are sol\-vab\-le by a simple iteration procedure}. At each step, we obtain an integral equation of the type
\begin{equation}\label{SS_1}
g_\alpha  (p) = \sum\limits_{\alpha '} {\int\limits_B {\rd^3 p'} K_{\alpha \alpha '}^N } \left(p,p'\right)h_{\alpha '} \left(p'\right)
\end{equation}
for a function $h_\alpha(p)$ with additional conditions that eliminate arbitrariness in $h_\alpha(p)$ of the form $c_n p_n+c\,\varepsilon_{\alpha p}$ [$g_\alpha(p)$ is a known function]. Values $\nu, \nu_T$, and $h$ in this equations and kinetic coefficients of the system are calculated using the formulas (\ref{S_27}). The mentioned arbitrariness is related to conservation of energy and momentum in normal phonon processes. {\it Further analysis of equations of the type} (\ref{SS_1}) {\it requires additional information on the spectrum $\varepsilon _ {\alpha p}$ of the phonons and amplitude of their interaction $\Phi(\alpha_1 p_1;\alpha_2 p_2,\alpha_3 p_3)$}. It should be also stressed that necessary solutions of equation (\ref{SS_1}) are  vectors and tensors of second rank depending on momentum. Even for crystals of the considered cubic symmetry classes $O$, $O_h$, $T_h$, they have a rather complicated structure.

Note also that at low temperatures, the long-wavelength acoustic phonons give the main contribution to thermodynamic and kinetic properties of a crystal (see, for example, \cite{Sok_6}). This allows one to find phonon spectrum $\varepsilon_{ap}$ and phonon interaction amplitude  $\Phi(\alpha_1 p_1; \alpha_2 p_2,\alpha_3 p_3)$ based on the elasticity theory. This leads to some simplification of the above obtained equations that will be analyzed elsewhere (the standard theory in this approach has been constructed in \cite{Sok_6, Sok_7}).

Let us present some results of our calculations in the main low temperature approximation:
\[
A_ {\alpha n} ^ {[0]} (p) = \frac {3\sigma} {\left\langle p^2\right\rangle} p_n\,, \qquad
\nu ^{[1]} = \frac{\sigma} {\left\langle p^2\right\rangle} \left\{p_n, p_n \right\} ^ U, \qquad h^{[0]}= \frac{3\sigma T^2}{\left\langle p^2 \right\rangle}\frac{\partial p}{\partial T}\,;
\]
\[
\nu _T ^ {[1]} = \frac {1}{{3T}}\left\{p_n,C_{\alpha n}^{[0]}(p)\right\}^U, \qquad
\kappa ^ {[0]} = \frac {{T ^ 2}} {3} \left\{C_ {\alpha n} ^ {[0]} (p), C_{\alpha n}^ {[0]}(p)\right\}^N,
\]
\begin{equation}\label{S_39}
\eta _ {nl, ms} ^ {[0]} = \frac{\left\langle p^2\right\rangle}{3\sigma} \left\{D_ {\alpha nl} ^ {[0]} (p), D_ {\alpha ms} ^ {[0]} (p) \right\}^N.
\end{equation}
Among other things, these formulas give positively defined expressions for the heat conductivity and viscosity of the system. Some results of the main approximation coincide with the corresponding results of the standard theory
\[
A_ {\alpha n}^{[0]}(p) = A_ {\alpha n}^o(p), \qquad
B_ {\alpha nl}^{[0]}(p) = B_{\alpha nl}^o(p);
\]
\begin{equation}\label{S_56}
\nu^{[1]} = \nu^o,
\qquad h^{[0]} = h^o, \qquad \mu _ {nl, ms}^{[0]} = \mu_{nl, ms}^o\,,\qquad (\text{at  \,}\, \sigma=\sigma^o, \,\,\, a=a^o).
\end{equation}
Functions $C_ {\alpha n} ^ o$, $D_ {\alpha n \, l}^o$ satisfy the integral equations (\ref{S_21}), (\ref{S_22}) with quantities $A_ {\alpha n}^o, \sigma^o, h^o, \nu^o, \nu_T^o$ taken from \eqref{S_33}--\eqref{S_34} instead of $A_ {\alpha n}, \sigma, h, \nu, \nu_T$. These equations can be solved at low temperatures in the perturbation theory in $\mu$ that gives the following results
\[
C_ {\alpha n} ^ {[0]}(p) = C_ {\alpha n} ^ {o[0]}(p), \qquad
D_ {\alpha nl} ^ {[0]}(p) = D_ {\alpha n \, l} ^ {o[0]}(p);
\]
\begin{equation}\label{S_57}
\kappa_T ^ {[1]} = \kappa_T^ {o[1]}, \qquad \kappa ^ {[0]} = \kappa ^ {o[0]}, \qquad  \eta _ {nl, ms}^{[0]} = \eta_{nl, ms}^{o[0]}\,,
\qquad (\text{at  \,}\, \sigma=\sigma^o, \,\,\, a=a^o)
\end{equation}
(notation like $A^{o[s]}$ gives a contribution of the order $\mu^s$ to a quantity of the standard theory $A^o$). So, {\it results of the developed theory and the standard theory coincide with one another in the main low temperature approximation} (at $\sigma=\sigma^o, a=a^o$, i.e., at the standard definition of the drift velocity and temperature).

The first two formulas (\ref{S_56}) show that at low temperatures, the difference between the phonon distribution function in the zeroth approximation in gradients ${\rm {f}} _ {\alpha p} ^ {(0)} (x, \xi)$ and the local Planck distribution $n_ {\alpha p} (\xi (x))$ is estimated by ${\rm {f}} _{\alpha p}^{(0)}(x,\xi) = n_{\alpha p}(\xi (x))+O(\lambda^1\mu^1, \lambda^2\mu^1, \lambda^3)$. Therefore, {\it at low temperatures $T \ll T_{\mathrm{D}}$ and small drift velocity, the violation of the local equilibrium in the phonon hydrodynamics is exponentially small}. This result confirms the applicability  of the standard theory at low temperatures.

\section{The steady states of insulators \label{E}}

Investigation of steady states of the system is of great interest because their properties are easier to be analyzed experimentally. The hydrodynamic equation for the drift velocity (\ref{S_31}) should be solved in the steady state with respect to the drift velocity $u_n$ in the form of a series in temperature gradients  that gives
\begin{equation}\label{S_61}
u_n = - \frac {1} {\nu} \left ({\nu _T + \frac {{\partial p}} {{\partial T}}} \right) \frac {{\partial T}} { {\partial x_n}} + O \left(g ^ 3\right).
\end{equation}
The accuracy of this result is limited by the accuracy estimation in (\ref{S_31}). It should be noted that an equation of the type (\ref{S_61}) appears in the standard theory of steady processes in dielectrics too [see, for example, \cite{Sok_4} where (\ref{S_61}) is obtained for $T\ll T_0$ without a term with $\nu_T$ that is exponentially small according to (\ref{S_37})]. Taking into account the result (\ref{S_61}), the expression (\ref{S_26}) for phonon energy flux yields
\begin{equation}\label{S_62}
q_n = - \tilde \kappa \frac {{\partial T}} {{\partial x_n}} + O \left(g ^ 3\right),\qquad  \tilde \kappa \equiv \kappa + \frac {h} {\nu} \left(\nu _T + \frac {{\partial p}}{{\partial T}}\right)
\end{equation}
where we have introduced thermal conductivity in a steady state $\tilde \kappa$. The formula shows that in the absence of the Umklapp processes, conductivity $\tilde \kappa = \infty$ [according to (\ref{S_27}), in this case $\nu = 0$, $\nu _T = 0$]. {\it This well-known result is not surprising since an isolated system does not have nonequilibrium steady states}. {\it Only the presence of Umklapp processes that lead to a nonconservation of the phonon momentum of a dielectric crystal makes steady states possible}.

There is an interesting issue concerning the conditions of stationarity of the phonon temperature of such a system. Substitution of the expression for the drift velocity in a steady state (\ref{S_61}) into equation (\ref{S_31}) for temperature, yields a condition
\begin{equation}\label{S_65}
\frac {\partial} {{\partial x_n}} \left ({\tilde \kappa \frac {{\partial T}} {{\partial x_n}}} \right) = 0
\end{equation}
(in the second order of the perturbation theory in gradients). The meaning of this result as the condition of heat source absence in a steady state is clear. At the same time, the issue regarding the temperature distribution in isolated insulators is not discussed in the literature, although the nature of steady states in such systems is quite unusual.

Steady state thermal conductivity of the phonon system of an insulator $\tilde \kappa$ can be calculated at low temperatures based on the results of the previous section that in the main approximation gives
\begin{equation}\label{S_66}
\tilde \kappa = \tilde \kappa ^ {[- 1]}+ O \left(\mu^0\right), \qquad \tilde \kappa^ {[- 1]} = \frac {3T^2}{\left\{p_n, p_n\right\}^U}\left ({\frac {\partial p} {\partial T}} \right)^ 2.
\end{equation}
{\it This result coincides with the result by Akhiezer} \cite{Sok_2} and does not depend on the definition of the drift velocity (\ref{S_14}). The thermal conductivity $\tilde\kappa^{[- 1]}$  is exponentially large which  eliminates the problem of calculating the corrections to it at $T \ll T_{\mathrm{D}}$.

\section{Conclusions}

In this paper, the Chapman-Enskog method is generalized to take into account relaxation processes (kinetic modes) in the hydrodynamic theory, i.e., the processes that can be present in a spatially homogeneous state of a system. On this basis:
\begin{itemize}
\item Nonlinear hydrodynamics of the phonon system has been built in the approximation of small phonon drift velocity for insulators with cubic symmetry of the lattice .
\item  It is proved that a small phonon drift velocity decays.
\item  In the perturbation theory in the drift velocity it is found, that in the phonon hydrodynamics the distribution function of phonons in the zeroth approximation in the gradients is different from the Planck distribution with the velocity. At the same time, at zero drift velocity, these distributions coincide. So, it is established that the local equilibrium in the phonon hydrodynamics is violated.
\item It is shown, that at low temperatures the integral equations of the theory are solvable iteratively.
\item  It is shown, that at low temperatures and at small drift velocity, the local equilibrium in the hydrodynamic phonon system takes place with exponential accuracy.
\item  The obtained results can be applied to the analysis of hydrodynamic processes in the system of phonons of the insulator at intermediate and high temperatures.
\item ~The relation between usual hydrodynamic thermal conductivity and thermal conductivity of phonons in the steady state has been established. It is shown that the Akhiezer expression gives the main low temperature contribution to the thermal conductivity of an insulator in its steady states.
\item  The thermal conductivity of an insulator is impossible to be calculated without taking into account the Umklapp processes because nonequilibrium steady states in a closed system do not exist without them.
\end{itemize}

\section*{Acknowledgements}
This work was supported in part by the State Foundation for Basic Research of Ukraine (project No. 25.2/102).

\ukrainianpart

\title{Гідродинамічні стани фононів діелектриків}

\author{С.О. Соколовський}

\address{Придніпровська державна академія будівництва та архітектури}
\makeukrtitle
\begin{abstract}
\tolerance=3000%
Метод Чепмена-Енскога узагальнено для врахування впливу кінетичних мод системи на гідродинамічну еволюцію. У наближенні малої дрейфової швидкості вивчено гідродинамічні стани фононної підсистеми діелектрикa. Для спрощення, дослідження проведено для кристалів кубічних класів симетрії. Встановлено, що у фононній гідродинаміці локальна рівновага порушується навіть у лінійному наближенні за швидкістю. Це є наслідком відсутності закону збереження імпульсу, що веде до релаксації дрейфової швидкості. Отримано рівняння фононної гідродинаміки з урахуванням дисипативних процесів. Результати порівняно зі стандартною теорією, яка базується на наявності локальної рівноваги. Отримано інтегральні рівняння для розрахунку об'єктів теорії (включаючи в'язкість та теплопровідність). Показано, що у границі низьких температур ці рівняння розв'язуються ітераціями. Розглянуто стаціонарні стани системи і отримано вираз для її стаціонарної теплопровідності. Показано, що вона у низькотемпературній границі співпадає з відомим результатом Ахієзера. Встановлено, що розподіл температур у стаціонарному стані діелектрика задовольняє умову відсутності джерела тепла.
\keywords фонони діелектрика, процеси перекидання, релаксаційні ступені вільності, локальна рівновага, метод Чепмена-Енскога, фононна гідродинаміка, мала дрейфова швидкість, низькі температури, стаціонарні стани
\end{abstract}


\newpage
\begin{thebibliography}{20}
\bibitem{Sok_1}
Peierls~R., Ann. Phys., 1929, \textbf{395}, No.~8, 1055; \doi{10.1002/andp.19293950803}.
\bibitem{Sok_2}
Akhiezer~A., Sov. Phys. JETP, 1940, \textbf{10}, No.~12, 1354.
\bibitem{Sok_4}
Lifshitz~E.M., Pitaevskii L.P., Physical Kinetics. Pergamon Press, Oxford, 1981.
\bibitem{Sok_5}
Gurevich~V.L., Transport in Phonon Systems. North-Holland, Amsterdam, 1988.
\bibitem{SSok_3}
Gurzhi~R.N., Sov. Phys. Usp., 1968, \textbf{11}, 255; \doi{10.1070/PU1968v011n02ABEH003815}.
\bibitem{Sok_6}
Akhiezer~A., Alexin~V., Khodusov~V., Low Temp. Phys., 1994, \textbf{20}, No.~12, 939; \doi{10.1063/1.592848 }.
\bibitem{Sok_7}
Akhiezer~A., Alexin~V., Khodusov~V., Low Temp. Phys.,1995, \textbf{21}, No.~1, 1; \doi{10.1063/1.593042}.
\bibitem{SSok_2}
Batsevych~O.F., Mryglod~I.M., Rudavskii~Yu.K., Tokarchuk~M.V., J. Phys. Stud., 2003, \textbf{7}, No.~3, 291 (in Ukrainian).
\bibitem{SSok_1}
Mryglod~I.M., Condens. Matter Phys., 1998, \textbf{1}, No.~4(16), 753.
\bibitem{SSok_4}
Sokolovsky~S.A., Theor. Math. Phys., 2011, \textbf{168}, No.~2, 1150; \doi{10.1007/s11232-011-0093-z}.
\bibitem{Sok_3}
Akhiezer~A.I., Peletminsky~S.V., Methods of Statistical Physics. Pergamon Press, Oxford, 1981.
\bibitem{SSok_5}
Garanin~D.A., Lutovinov~V.S., Ann. Phys., 1992, \textbf{218}, 293; \doi{10.1016/0003-4916(92)90089-5}.
\bibitem{Sok_8}
Schouten~J.A., Tensor Analysis for Physicists. Dover Publications, New York, 2011.
\end{thebibliography}
\end{document}